\documentclass[twocolumn,showpacs,aps,prl,superscriptaddress]{revtex4}

\usepackage{graphicx}
\usepackage{dcolumn}
\usepackage{amsmath}
\usepackage{epsfig}
\usepackage{verbatim}
\usepackage{multirow}
\usepackage{xspace}
\usepackage{relsize}

\def\babar{\mbox{\slshape B\kern-0.1em{\smaller A}\kern-0.1em
    B\kern-0.1em{\smaller A\kern-0.2em R}}}

\def\epem       {\ensuremath{e^+e^-}\xspace}
\def\tautau     {\ensuremath{\tau^+\tau^-}\xspace}

\def\ccbar {\ensuremath{c\overline c}\xspace}
\newcommand{\uds}      {\ensuremath{uds}}
\def\BR         {{\ensuremath{\cal B}\xspace}}
\def\DeltaE     {\mbox{$\Delta E$}\xspace}

\newcommand{\gev}{\ensuremath{\mathrm{\,Ge\kern -0.1em V}}\xspace}
\newcommand{\gevcc}{\ensuremath{{\mathrm{\,Ge\kern -0.1em V\!/}c^2}}\xspace}
\newcommand{\mevcc}{\ensuremath{{\mathrm{\,Me\kern -0.1em V\!/}c^2}}\xspace}

\def\invfb   {\ensuremath{\mbox{\,fb}^{-1}}\xspace}

\def\L{{\ensuremath{\cal L}}\xspace}
\def\pep2{PEP-II}

\newcommand{\jprlBase}       {Phys.\ Rev.\ Lett.\xspace}
\newcommand{\jprBase}        {Phys.\ Rev.\xspace}
\newcommand{\jplBase}        {Phys.\ Lett.\xspace}
\newcommand{\nimBaseA}       {Nucl.\ Instrum.\ Methods Phys.\ Res., Sect.\ A\xspace}
\newcommand{\npBase}         {Nucl.\ Phys.\xspace}
\newcommand{\nima}      [1]  {\nimBaseA~{\bf #1}}
\newcommand{\npb}       [1]  {\npBase\ B~{\bf #1}}
\newcommand{\plb}       [1]  {\jplBase\ B~{\bf #1}}
\newcommand{\jprl}      [1]  {\jprlBase\ {\bf #1}}
\newcommand{\jprd}      [1]  {\jprBase\ D~{\bf #1}}

\newcommand{\taulV }  {\ensuremath{\tau^{-}\!\to\ell^- V^0}}
\newcommand{\dMdE}    {\ensuremath{(\Delta M, \Delta E)}}
\newcommand{\deltaM}  {\ensuremath{\Delta M}}
\newcommand{\deltaE}  {\ensuremath{\Delta E}}

\newcommand{\EPHI}      {\ensuremath{e \phi }}
\newcommand{\MPHI}      {\ensuremath{\mu \phi }}
\newcommand{\ERHO}      {\ensuremath{e \rho }}
\newcommand{\MRHO}      {\ensuremath{\mu \rho }}
\newcommand{\EKS}       {\ensuremath{e K^* }}

\newcommand{\MKS}       {\ensuremath{\mu K^* }}

\newcommand{\EKSB}      {\ensuremath{e \overline{K}^* }}

\newcommand{\MKSB}      {\ensuremath{\mu \overline{K}^* }}

\newcommand{\taumphi}   {\ensuremath{\tau^- \to \mu^- \phi}}
\newcommand{\tauerho}   {\ensuremath{\tau^- \to e^- \rho}}
\newcommand{\taumrho}   {\ensuremath{\tau^- \to \mu^- \rho}}

\newcommand{\taumks}    {\ensuremath{\tau^- \to \mu^- K^*}}

\def\kk2f       {\mbox{\tt KK2f}\xspace}
\def\tauola     {\mbox{\tt TAUOLA}\xspace}
\def\photos     {\mbox{\tt PHOTOS}\xspace}
\newcommand{\Nobs}      {\ensuremath{N_{\rm obs}}}
\newcommand{\Nbgd}      {\ensuremath{N_{\rm bgd}}}
\newcommand{\Nul}       {\ensuremath{N^{90}_{\rm{UL}}}}
\newcommand{\BRul}      {\ensuremath{\BR^{90}_{\rm{UL}}}}
\newcommand{\BRulexp}   {\ensuremath{\BR^{90}_{\rm{exp}}}}

\newcommand{\lumi}    {451\invfb}
\newcommand{\lumiOn}  {410\invfb}
\newcommand{\lumiOff} {40.8\invfb}

\newcommand{\BABARPubYear}    {09}
\newcommand{\BABARPubNumber}  {002}
\newcommand{\SLACPubNumber}   {13559}

\begin{document}

\preprint{\babar-PUB-\BABARPubYear/\BABARPubNumber} 
\preprint{SLAC-PUB-\SLACPubNumber} 

\begin{flushleft}
\babar-PUB-\BABARPubYear/\BABARPubNumber\\
SLAC-PUB-\SLACPubNumber\\
\end{flushleft}

\title{
	{\large \bf \boldmath
	Improved Limits on Lepton Flavor Violating Tau Decays to $\ell\phi,\ell\rho,\ell K^*$, and $\ell\overline{K}^*$}
}

%% author list as of 09-Jan-2009 (489 authors)
%
\author{B.~Aubert}
\author{Y.~Karyotakis}
\author{J.~P.~Lees}
\author{V.~Poireau}
\author{E.~Prencipe}
\author{X.~Prudent}
\author{V.~Tisserand}
\affiliation{Laboratoire d'Annecy-le-Vieux de Physique des Particules (LAPP), Universit\'e de Savoie, CNRS/IN2P3, F-74941 Annecy-Le-Vieux, France }
\author{J.~Garra~Tico}
\author{E.~Grauges}
\affiliation{Universitat de Barcelona, Facultat de Fisica, Departament ECM, E-08028 Barcelona, Spain }
\author{M.~Martinelli$^{ab}$}
\author{A.~Palano$^{ab}$ }
\author{M.~Pappagallo$^{ab}$ }
\affiliation{INFN Sezione di Bari$^{a}$; Dipartimento di Fisica, Universit\`a di Bari$^{b}$, I-70126 Bari, Italy }
% ...
\collaboration{The \babar\ Collaboration}
\noaffiliation

\begin{abstract}
We search for the neutrinoless, lepton-flavor-violating tau decays
\taulV{}, where $\ell$ is an electron or muon and $V^0$ is a vector meson
reconstructed as $\phi\to K^+K^-,\rho\to\pi^+\pi^-,
K^*\to K^+\pi^-,\overline{K}^*\to K^-\pi^+$.
The analysis has been performed using \lumi\ of data collected at an \epem\
center-of-mass energy near 10.58\gev with the \babar\ detector 
at the \pep2\ storage rings. The number of events found 
in the data is compatible with the background expectation, 
and upper limits on the branching fractions are set in the range 
$(2.6-19) \times10^{-8}$ at the 90\% confidence level.
\end{abstract}

\maketitle

%%
%% --------- Introduction ----------------
%%
Lepton-flavor violation (LFV) involving tau leptons has 
never been observed, and recent experimental results have placed 
stringent limits on the branching fractions for 2- and 3-body 
neutrinoless tau decays~\cite{taulll, taulhh, belle08}.
Many descriptions of physics beyond the Standard Model (SM) 
predict such decays~\cite{paradisi05, brignole03};
and certain models~\cite{brignole04,arganda08} 
specifically predict semileptonic tau decays 
such as $\tau\to\ell~\phi/\rho/K^*/\overline{K}^*$ (\taulV), 
with rates as high as the current experimental limits~\cite{belle08}. 
An observation of these decays would be a 
clear signature of physics beyond the SM, while improved 
limits will further constrain models of new physics.

%%
%% --------- Data ----------------
%%
This paper presents a search for LFV in a set of eight neutrinoless
decay modes \taulV\cite{cc}, where $\ell$ is an electron or muon
and $V^0$ is a neutral vector meson decaying to two charged hadrons
($V^0\to h^+h^-$)
via one of the following four decay modes: 
$\phi\to K^+K^-$,
$\rho\to\pi^+\pi^-$,
$K^*\to K^+\pi^-$,
$\overline{K}^*\to\pi^+K^-$.
This analysis is based on data recorded 
by the \babar\ detector at the \pep2\ asymmetric-energy \epem\ 
storage rings operated at the SLAC National Accelerator Laboratory.
The \babar\ detector is described in detail in Ref.~\cite{detector}.
The data sample consists of \lumiOn\ recorded at an
\epem\ center-of-mass (c.m.) energy 
$\sqrt{s} = 10.58 \gev$, and \lumiOff\ recorded at
$\sqrt{s} = 10.54 \gev$.
With a calculated cross section for tau pairs 
of $\sigma_{\tau\tau} = 0.919\pm0.003$ nb \cite{tautau,kk}
at the stated luminosity-weighted $\sqrt{s}$,
this data set corresponds to the production of about 830 million tau decays.

%%
%%--------- Simulation ---------------
%%
We use a Monte Carlo (MC) simulation of lepton-flavor-violating 
tau decays to optimize the search.
Tau-pair events including higher-order radiative
corrections are generated using \kk2f \cite{kk}.
One tau decays via two-body phase space to a lepton and a vector meson, 
with the meson decaying according to the measured branching fractions \cite{PDG}. 
The other tau decays via SM processes simulated with \tauola \cite{tauola}.
Final state radiative effects are simulated for all decays 
using \photos \cite{photos}.
The detector response is modeled with \mbox{\tt GEANT4}~\cite{geant},
and the simulated events are then reconstructed in the same 
manner as data. SM background processes are modeled with a similar
software framework. 

%%
%% --------- Pre-Selection ----------------
%%
We search for the signal decay $\tau^-\to\ell^-V^0\to\ell^-h^+h^-$ by
reconstructing $\epem\to\tautau$ candidates
in which three charged particles, each identified 
as the appropriate lepton or hadron, 
have an invariant mass and energy close to that of the parent tau lepton. 
Candidate signal events are first required 
to have a ``3-1 topology,'' where one tau decay yields three
charged particles, while the second tau decay yields one charged particle.
This requirement on the second tau decay greatly reduces the background
from continuum multi-hadron events. 
Events with four well-reconstructed tracks and zero net charge are selected,
and the tracks are required to point toward a common region consistent with 
\tautau production and decay. 
The polar angle of all four tracks in the laboratory
frame is required to be within the calorimeter acceptance.
Pairs of oppositely-charged tracks are ignored 
if their invariant mass, assuming electron mass hypotheses, 
is less than 30\mevcc. Such tracks are
likely to be from photon conversions in the traversed material.
The event is divided into hemispheres 
in the \epem\ c.m. frame using the plane perpendicular to the thrust axis,
as calculated from the observed tracks and neutral energy deposits.
The signal (3-prong) hemisphere must contain exactly three tracks 
while the other (1-prong) hemisphere must contain exactly one.
Each of the charged particles found in the 3-prong 
hemisphere must be identified as a lepton or hadron
candidate appropriate to the search channel. 
The relevant particle identification capabilities of the \babar\
detector are described in Ref.~\cite{taulhh}.

%%
%% ------------ Selection ----------------
%%
To further suppress backgrounds from quark pair production, 
Bhabha scattering events, and SM tau pair production, 
we apply additional selection criteria 
separately in the eight different search channels. 
Specific cut values are shown in Tab.~\ref{tab:cuts}.
All selection criteria are optimized 
to provide the smallest expected upper limit on the branching fraction 
in the background-only hypothesis. 
Resonant decays are selected with cuts on the invariant mass 
of the two hadrons in the 3-prong hemisphere ($m_{hh}$).
The invariant mass of the 1-prong hemisphere ($m_{1-pr}$) is
calculated from the charged and neutral particles 
in that hemisphere and the total missing momentum in the event.
As the missing momentum in signal events results from 
one or more neutrinos in the 1-prong hemisphere,
this mass is required to be near the tau mass.
Background events from quark pair production 
are suppressed with cuts on 
the missing transverse momentum in the event ($p_T^{miss}$),
the scalar sum of all transverse momenta in the c.m.~frame ($p_T^{cms}$),
and the number of photons in the 1-prong and 3-prong hemispheres 
($n^\gamma_{1pr},n^\gamma_{3pr}$).
To reduce the background contribution from radiative Bhabha and di-muon events, 
the 1-prong and 3-prong momentum vectors
must not be collinear in the c.m. frame. 
For the same reason, the 1-prong track must not be identified 
as an electron for the \tauerho\ search. 

\begin{table}[htb]
\caption{Values of the cuts on the selection variables described in the text. 
Masses are in units of GeV/c$^2$, and momenta in units of GeV/c.}
\begin{center}
\begin{tabular}{l|cccccccc}
\hline\hline\vspace{-3.0mm}\\
Channel               & \EPHI\ &\MPHI\ &\ERHO\ & \MRHO\ & \EKS\ &\MKS\ &\EKSB\ &\MKSB\ \\
\hline
$m_{hh}$ min          & 1.000  & 1.005 & 0.6   & 0.6    & 0.8   & 0.82 & 0.80  & 0.78\\
$m_{hh}$ max          & 1.040  & 1.035 & 0.92  & 0.96   & 1.0   & 0.98 & 1.04  & 1.00\\
$m_{1-pr}$ min        & 0.3    & 0.4   & 0.3   & 0.3    & 0.3   & 0.2  & 0.3   & - \\
$m_{1-pr}$ max        & 2.5    & 2.5   & 2.5   & 2.5    & 2.5   & 2.5  & 2.5   & - \\
$p_T^{miss}$ min      & 0.4    & 0.3   & 0.4   & 0.4    & 0.4   & 0.4  & 0.4   & 0.4 \\
$p_T^{cms}$ min       & 0.5    & -     & -     & -      & 0.6   & -    & 0.3   & - \\
$n^\gamma_{1pr}$ max & 4      & 3     & 3     & 1      & -     & 3    & -     & 2 \\
$n^\gamma_{3pr}$ max & 3      & 1     & 2     & 1      & -     & 2    & -     & 1 \\
\hline\hline
\end{tabular}
\end{center}
\label{tab:cuts}
\end{table}

%%
%% ----------- deltaM/deltaE ------------
%%
As a final discriminant, we require candidate signal events to have
an invariant mass and total energy in the 3-prong
hemisphere consistent with a parent tau lepton.
These quantities are calculated from the measured track momenta, 
assuming lepton and hadron masses that correspond to the neutrinoless tau decay 
in each search channel.
The energy difference is defined as 
$\Delta E \equiv E^{\star}_{\mathrm{rec}} - E^{\star}_{\mathrm{beam}}$,
where $E^{\star}_{\mathrm{rec}}$ is the total energy of the tracks
observed in the 3-prong hemisphere and $E^{\star}_{\mathrm{beam}}$
is the beam energy, with both quantities calculated in the c.m. frame.
The mass difference is defined as
$\Delta M \equiv M_{\mathrm{EC}} - m_{\tau}$ where $M_{\mathrm{EC}}$ 
is calculated from a kinematic fit to the 3-prong track momenta with
the energy constrained to be $\sqrt{s/2}$ in the c.m. frame,
and $m_{\tau}=1.777\gevcc$ is the tau mass \cite{PDG}. 
While the energy constraint significantly reduces 
the spread of \deltaM\ values,
it also introduces a correlation between
\deltaM\ and \deltaE{}, which must be taken into account 
when fitting distributions in this 2-dimensional space.

\begin{figure*}[!htbp]
\begin{center}
\resizebox{0.99\textwidth}{!}{
\includegraphics{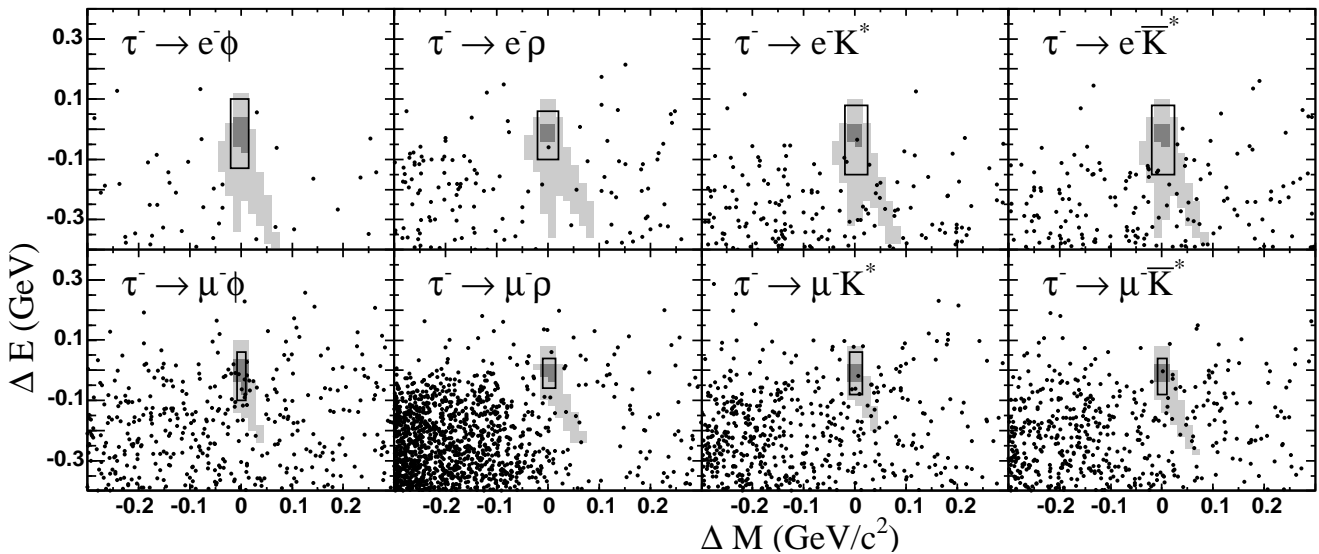}
}
\caption{Observed data shown as dots in the Large Box of the \dMdE\ plane and 
the boundaries of the Signal Box.
The dark and light shading indicates contours containing
50\% and 90\% of the selected MC signal events, respectively.}
\label{fig1}
\end{center}
\end{figure*}

Detector resolution and radiative effects broaden
the signal distributions in the \dMdE\ plane. 
Because of the correlation between \deltaM\ and \deltaE{},
the radiation of photons from the incoming \epem\ particles
produces a tail at positive values of \deltaM\ 
and negative values of \deltaE{}.
Radiation from the final-state leptons, 
which is more likely for electrons than muons, 
leads to a tail at low values of $\Delta E$.
Rectangular signal boxes (SB) in the \dMdE\ plane are defined separately 
for each search channel.
As with previous selection criteria, the SB boundaries are chosen
to provide the smallest expected upper limit on the branching fraction.
The expected upper limit is estimated using only MC
simulations and data events in the sideband region, 
as described below. 
Figure~\ref{fig1} shows the observed data in the Large Box (LB) of the \dMdE\ plane, 
along with the SB boundaries 
and the expected signal distributions.
Table~\ref{tab:box} lists the channel-specific dimensions of the SB. 
While a small fraction of the signal events lie outside the SB,
the effect on the final result is negligible. 
To avoid bias, we use a blinded analysis procedure
with the number of data events in the SB
remaining unknown until the selection criteria 
are finalized and all crosschecks are performed.

\begin{table}[htb]
\begin{center}
\caption{Signal Box boundaries; 
$\Delta M$ is in units of \gevcc\ and $\Delta E$ in units of \gev{}.}
\begin{tabular}{l|cccccccc}
\hline\hline\vspace{-3.0mm}\\
Mode       &\EPHI\ &\ERHO\ &\EKS\  &\EKSB\  &\MPHI\  &\MRHO\ &\MKS\  &\MKSB\  \\
\hline
$\Delta M_{\rm min}$ & -0.02 & -0.02 & -0.02 & -0.015 & -0.008 & -0.01 & -0.01 & -0.008 \\
$\Delta M_{\rm max}$ & 0.015 & 0.02  & 0.02  & 0.02   & 0.01   & 0.015 & 0.01  & 0.01   \\
$\Delta E_{\rm min}$ & -0.13 & -0.10 & -0.15 & -0.125 & -0.09  & -0.06 & -0.08 & -0.08  \\
$\Delta E_{\rm max}$ & 0.10  & 0.06  & 0.08  & 0.06   & 0.06   & 0.04  & 0.04  & 0.06   \\
\hline\hline
\end{tabular}
\label{tab:box}
\end{center} 
\end{table}

%%
%% ------------- Backgrounds ---------
%%
There are three main classes of background events remaining after
the selection criteria are applied: charm quark production (\ccbar),
low-multiplicity continuum $\epem\to u\bar{u}/d\bar{d}/s\bar{s}$ events 
(\uds{}), and SM \tautau pair events.
The background from two-photon production is negligible. 
These three background classes have distinctive distributions
in the \dMdE\ plane.
The \uds\ events tend to populate the plane evenly,
with a fall-off at positive values of \DeltaE{}.
Events in the \ccbar\ sample exhibit peaks at positive
values of \deltaM\ due to $D$ and $D_s$ mesons, 
and are generally restricted to negative values
of \deltaE{}. The \tautau background events
are restricted to negative values of both $\Delta E$ and $\Delta M$.

The expected background rates in the SB are determined by
fitting a set of 2-dimensional probability density functions (PDFs) to the
observed data in the grand sideband (GS) region of the \dMdE\ plane.
The GS region is defined as the LB minus the SB. 
The shapes of the PDFs are determined by fits to the \dMdE\ distributions 
of background MC samples in the LB, as described in Ref.~\cite{taulll}.
The present analysis makes use of the same parameterization 
as Ref.~\cite{taulll} for the \deltaE\ spectra, 
except for the case of the \ccbar\ spectrum in some search channels.
In these cases, combinations of polynomial and Gaussian functions are used.
The choice of PDF for the \deltaM\ spectrum of the \uds\ samples is 
the same as used in Ref.~\cite{taulll}, while the \tautau\ and \ccbar\
\deltaM\ spectra are modeled with Gaussian and polynomial functions, 
or the Crystal Ball function~\cite{CBF}.
All shape parameters, including a rotation angle accounting for the
correlation between \deltaE\ and \deltaM{}, 
are determined from the fits to MC samples. 

Once the shapes of the three background PDFs are determined, 
an unbinned extended maximum likelihood fit to the data in the GS region
is used to find the expected background count in the SB.
The fits to the background MC samples and to data are performed
separately for each of the eight search channels.

\renewcommand{\multirowsetup}{\centering}

%%
%%---------- Signal efficiency ---------
%%
We estimate the signal event selection efficiency 
with a MC simulation of lepton-flavor violating tau decays.
Between $20\%$ and $40\%$ of the MC signal events pass the 3-1 topology requirement.
The efficiency for identification of the three final-state particles ranges 
from $42\%$ for \taumks{} to $82\%$ for \tauerho{}.
The total efficiency for signal events to be found
in the SB is shown in Table~\ref{tab:results},
and ranges from 4.1\% to 8.0\%.
This efficiency includes the branching fraction for the vector meson decay
to charged daughters, 
as well as the branching fraction for 1-prong tau decays.

%%
%%---------- Systematics ---------
%%
The particle identification efficiencies and misidentification probabilities
have been measured with control samples both for data and 
MC events, as a function of particle momentum, polar angle, 
and azimuthal angle in the laboratory frame. 
The systematic uncertainties related to the particle identification have
been estimated from the statistical uncertainty of the
efficiency measurements and from the difference between data and
MC efficiencies. These uncertainties range from 1.7\% for \tauerho\ 
to 9.0\% for \taumrho{} \cite{uncertain}.
The modeling of the tracking efficiency and the uncertainty from the 1-prong tau
branching fraction each contribute an additional 1\% uncertainty. 
Furthermore, the uncertainty on the intermediate branching
fractions $\mathcal{B}(\phi,K^*,\overline{K}^*\to h^+h^-)$ contributes a $1\%$ uncertainty.
All other sources of uncertainty in the signal efficiency are found to be 
negligible, including the statistical limitations of the MC signal samples,
modeling of radiative effects by the generator,
track momentum resolution, trigger performance,
and the choice of observables used for event selection.

Since the background levels are extracted directly from the data,
systematic uncertainties on the background estimation are directly
related to the background parameterization and the fit technique used.
Uncertainties related to the fits to the background samples are estimated 
by varying the background shape parameters according to the covariance matrix
and repeating the fits, and range from $3.8\%$ to $10\%$.
Uncertainties related to the fits for the background yields in the GS
are estimated by varying the yields within their errors,
and range from $4.1\%$ to $16\%$.
The total uncertainty on the background estimates 
is shown in Table~\ref{tab:results}.
Crosschecks of the background estimation are performed by 
comparing the number of events expected and observed in 
sideband regions immediately neighboring the SB for 
each search channel. No major discrepancies are observed.

\begin{table}
\begin{center}
\caption{Efficiency estimate, number of expected background events (\Nbgd),
number of observed events (\Nobs), 
observed upper limit at 90\% CL on the number of signal events (\Nul),
expected branching fraction upper limit at 90\% CL (\BRulexp),
and observed branching fraction upper limit at 90\% CL (\BRul). 
\BRulexp\ and \BRul\ are in units of $10^{-8}$.
}
\begin{tabular}{lccccccc}
\hline\hline\vspace{-3.0mm}\\
Mode & $\varepsilon$[\%] & ~\Nbgd  & \Nobs & ~\Nul & ~~\BRulexp & ~~\BRul  \\
\hline
\EPHI &$ 6.43 \pm 0.16 $&$ 0.68 \pm 0.12 $&$ 0 $&$ 1.8 $&$ 5.0 $&$ 3.1 $ \\
\MPHI &$ 5.18 \pm 0.27 $&$ 2.76 \pm 0.16 $&$ 6 $&$ 8.7 $&$ 8.2 $&$ 19  $ \\
\ERHO &$ 7.31 \pm 0.18 $&$ 1.32 \pm 0.17 $&$ 1 $&$ 3.1 $&$ 4.9 $&$ 4.6 $ \\
\MRHO &$ 4.52 \pm 0.41 $&$ 2.04 \pm 0.19 $&$ 0 $&$ 1.1 $&$ 8.9 $&$ 2.6 $ \\
\EKS  &$ 8.00 \pm 0.19 $&$ 1.65 \pm 0.23 $&$ 2 $&$ 4.3 $&$ 4.8 $&$ 5.9 $ \\
\MKS  &$ 4.57 \pm 0.36 $&$ 1.79 \pm 0.21 $&$ 4 $&$ 7.1 $&$ 8.5 $&$ 17  $ \\
\EKSB &$ 7.76 \pm 0.18 $&$ 2.76 \pm 0.28 $&$ 2 $&$ 3.2 $&$ 5.4 $&$ 4.6 $ \\
\MKSB &$ 4.11 \pm 0.32 $&$ 1.72 \pm 0.17 $&$ 1 $&$ 2.7 $&$ 9.3 $&$ 7.3 $ \\
\hline\hline
\end{tabular}
\label{tab:results}
\end{center}
\end{table}

%%
%%------------- Results -------
%%
The number of events observed (\Nobs) and the number of 
background events expected (\Nbgd) are shown in Table~\ref{tab:results}. 
The POLE calculator~\cite{pole},
based on the method of Feldman and Cousins~\cite{feldmanCousins},
is used to place $90\%$ CL upper limits on the number of signal events (\Nul), 
which include uncertainties on \Nbgd\ and on the selection efficiency ($\varepsilon$).
For the \taumphi\ search, the POLE calculation results 
in a two-sided interval at 90\% CL
for the number of signal events: $[0.39 - 8.65]$. 
Upper limits on the branching fractions are calculated according to 
$\BRul = \Nul/(2 \varepsilon \L \sigma_{\tau\tau})$, 
where the values $\L$ and $\sigma_{\tau\tau}$ are the
integrated luminosity and \tautau cross section, respectively.
The uncertainty on the product $\L \sigma_{\tau\tau}$ is 1.0\%. 
Table~\ref{tab:results} lists the upper limits on the branching fractions, 
as well as the expected upper limit \BRulexp{}, 
defined as the mean upper limit expected in the background-only hypothesis.
The 90\% CL upper limits on the $\tau\to\ell~\phi/\rho/K^*/\overline{K}^*$ branching fractions 
are in the range $(2.6-19)\times10^{-8}$,
and these limits represent improvements  
over the previous experimental bounds~\cite{belle08} in almost all search channels.

We are grateful for the excellent luminosity and machine conditions
provided by our \pep2\ colleagues, 
and for the substantial dedicated effort from
the computing organizations that support \babar.
The collaborating institutions wish to thank 
SLAC for its support and kind hospitality. 
This work is supported by
DOE
and NSF (USA),
NSERC (Canada),
CEA and
CNRS-IN2P3
(France),
BMBF and DFG
(Germany),
INFN (Italy),
FOM (The Netherlands),
NFR (Norway),
MES (Russia),
MEC (Spain), and
STFC (United Kingdom). 
Individuals have received support from the
Marie Curie EIF (European Union) and
the A.~P.~Sloan Foundation.


\begin{thebibliography}{99}
\bibitem{cc}
Throughout this paper, charge conjugate decay modes are implied.
\bibitem{uncertain}
All uncertainties quoted in the text are relative.
\end{thebibliography}
\end{document}